\documentclass[%
reprint, twocolumn,
superscriptaddress,
showpacs,preprintnumbers,
amsmath,amssymb, aps, pra,
]{revtex4}

\usepackage{graphicx}% Include figure files
\usepackage{dcolumn}% Align table columns on decimal point
\usepackage{bm}% bold math
\usepackage{color} % change color of text
\usepackage{CJK}
\usepackage{dsfont}
\usepackage[extension=xxx]{hyperref}
\usepackage{physics} %Useful to write bra-ket notation
\usepackage{soul}

\newcommand{\beq}{\begin{equation}}
\newcommand{\eeq}{\end{equation}}
\newcommand{\beqa}{\begin{eqnarray}}
\newcommand{\eeqa}{\end{eqnarray}}
\def\ra{\rangle}
\def\beq{\begin{equation}}

\definecolor{myorange}{RGB}{255,117,40}

\begin{document}
\title{Trapped-ion Fock state preparation by potential deformation}
\author{M. A. Simon}
\email{miguelangel.simon@ehu.eus}
\affiliation{Department  of Physical Chemistry, UPV/EHU, Apdo. 644, Bilbao 48080, Spain}
\author{M. Palmero}
\affiliation{Science and Math Cluster, Singapore University of Technology and Design, 8 Somapah Road, 487372 Singapore}
\author{S. Mart\'\i nez-Garaot}
\affiliation{Department  of Physical Chemistry, UPV/EHU, Apdo. 644, Bilbao 48080, Spain}
%\author{{\color{red}J. Alonso?}}
%\affiliation{Institute for Quantum Electronics, ETH Z\"urich, Otto-Stern-Weg 1, 8093 Z\"urich, Switzerland}
\author{J. G. Muga}
\affiliation{Department  of Physical Chemistry, UPV/EHU, Apdo. 644, Bilbao 48080, Spain}
%Añadir mas autores siguiendo la misma estructura
%\author{Author1}
%\email{email1}
%\affiliation{Afiliacion1}
%
\date{\today}
\begin{abstract}
We propose protocols to prepare highly excited energy eigenstates of a trapped ion in a harmonic trap which
do not require laser pulses to induce transitions among internal levels. Instead the protocols  rely on  smoothly deforming
the trapping potential between single and double well configurations. The speed of the changes
is set to minimize non-adiabatic transitions by keeping the adiabaticity parameter constant.  High fidelities are found for times more than two orders of magnitude smaller than with linear ramps of the control parameter. Deformation protocols are also devised to  prepare  superpositions to optimize interferometric sensitivity,
combining the ground state and a highly excited state. 
\end{abstract}
\pacs{37.10.Gh, 37.10.Vz, 03.75.Be}
%% PACS
%37.10.Gh Atom traps and guides
%37.10.Vz Mechanical effects of light on atoms, molecules, and ions
%03.75.Be Atom and neutron optics%
%32.80.Qk Coherent control of atomic interactions with photons
%32.80.Xx Level crossing and optical pumping
%33.80.Be Level crossing and optical pumping
%03.65.Yz Decoherence; %open systems; quantum statistical methods
%42.79.Fm ...beam splitters....
\maketitle
%
%
%%%%%%%%%%%%%%%%%%%%%%%%%%%%%%%%%%%%%%%%
%\begin{figure}[b!]
\begin{figure}
%\centering
  \includegraphics[width = 8.5 cm]{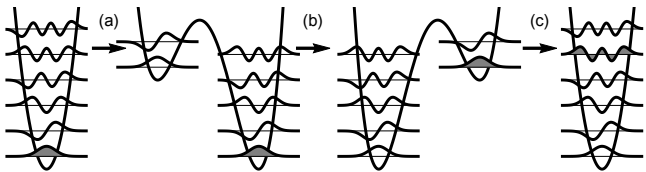}
  \caption{Scheme for Fock-state preparation. (a) Demultiplexing: splitting the harmonic trap into an asymmetrical double well; (b) Bias inversion of the double well; (c) Multiplexing: inverse of demultiplexing. The shaded wave functions are the ideal initial and target states.}
 %The level ordering at the start and the end of steps (a) and (c) is preserved so they could be performed adiabatically while in step (b) the level ordering is altered.}
 \label{fig:ProtocolDiagram}
\end{figure}
%%%%%%%%%%%%%%%%%%%%%%%%%%%%%%%%%%%%%%%%%%%%%
%
%\section{Introduction}
%\label{Section:Introduction}
%
%
%
{\it Introduction.}
%Trapped ions are among the most developed physical platforms for  quantum technologies \cite{Cirac1995,molmer1999multiparticle,BlattWinelandEntangledStatesTrappedIons},
A trapped-ion architecture for quantum technologies
rests on combining  basic operations such as
logic gates or shuttling, and generally needs controlling quickly and accurately internal and motional states.
Preparing (Fock) states with a well defined number of vibrational
quanta is one of the basic manipulations  that may be used to implement quantum memories,
entanglement operations, or  communications \cite{Martinez-Garaot2013,Galland2014}.  Fock states with large number of phonons
can be useful in metrology protocols based on \textit{NOON} states, which give measurement outcomes with
uncertainties reaching the Heisenberg bound \cite{Zhang2018,giovannetti2011advances}.
{Also, superpositions of eigenstates with maximally separated energies give optimal interferometric sensitivities \cite{Margolus1998, Caves2010} e.g. to measure motional frequency changes \cite{McCormick2019}.}
%A further motivation to control their formation
%is the possibility to encode information in vibrational
%modes  \cite{}.
%Vibrational Mode Multiplexing of Ultracold Atoms, see Fig. ....
Several  schemes to prepare Fock states have been proposed      \cite{Cirac1993,Cirac1994,Meekhof1996,davidovich1996quantum,DeMatos1996,abah2019quantum},
but for a trapped ion they have only been produced by sequences of Rabi pulses, which is in fact quite challenging,
as  an $n$-phonon state needs of the order of $n$ pulses with accurately defined frequency and area, so
%with a tight beam focused on a single ion.
the errors in intensity and
frequency, and timing imperfections reduce the fidelity \cite{Linington-Plenio-2008}.
{A recent experiment \cite{McCormick2019} applied such sequences of Rabi pulses with unprecedented accuracy to approximately reach Fock states of up to $n = 100$. To reach the highest Fock states,  higher order sidebands, i.e., pulses that jump more than a single level at a time (up to four in this case), had to be applied, but still the required time  and errors grow rapidly with the phonon number. }

The goal pursued here is to create an excited Fock state for a single ion from the ground state without laser-induced
internal transitions involved,  by means of deformations of the trap. The potentials in linear, multielectrode Paul ion traps can be deformed by programming
the voltages applied to the electrodes,
see  \cite{Kaufmann2014,Home2006,Nizamani2012,Furst2014}. Since these operations  only require that the trapped particle has an electrical charge they can be applied to other particles besides ions, like nano-particles \cite{Guan2011} or electrons \cite{Segal2006}. Operations
that do not use lasers to link internal and motional states
are worth exploring for quantum technologies since they would allow to create universal control devices independent
of the internal structure of the atom, and free from the usual disadvantages of laser control (frequency, position
and intensity instabilities, spontaneous decay) although, of course, they involve their own technical limitations and mass dependence. The present proposal
intends to demonstrate some possible benefits of the strategy based on trap deformations and
motivate further work to test and overcome these limitations.

The approach proposed here is depicted in
Fig. \ref{fig:ProtocolDiagram} and involves three steps:
(a) demultiplexing;
% the harmonic trap of an ion in its motional ground state is split into a biased double well trap so that the ion motional state finishes in the ground
%state of the lower well.}
(b) bias inversion;
%: the  bias of the double well is reversed in such a way that the ion stays
%in the same well. The level is still the ground state of that well, which becomes some excited state with respect to
%the double well.
(c) multiplexing.
%: the two wells are  merged back into a harmonic trap and the ion ends up in an excited Fock
%state of motion.
Steps (a) and (c) could be carried out adiabatically or using some shortcut to adiabaticity  (STA)  \cite{Torrontegui2013, Guery-Odelin2019}
since the level ordering at the start and at the end of the process
is conserved. For the  second step the ordering of the levels is altered, so  there is no global adiabatic mapping that connects
initial and final states. However, in a fast process the wells are effectively independent so that STA approaches
can also be applied as demonstrated in \cite{Martinez-Garaot2015}. A faster-than-adiabatic
approach for step (a) was applied  in \cite{Martinez-Garaot2013} with neutral atoms,
%, where the trapping potential is created by dipole interacti
but  only for a  two motional-level model. In this paper we design
step (c) using  an STA approach  to minimize the non-adiabatic transitions distributing them homogeneously along  the process time \cite{Martinez-Garaot2015a, Palmero2019}.
% (see Fig. \ref{fig:ProtocolDiagram} and more details in Fig. \ref{fig:DoubWellMulti}).
The first step requires a similar  protocol but in reverse.
{\it Multiplexing.}
Consider a single ion in a trap which is effectively one dimensional driven by the Hamiltonian
%. The motion in the longitudinal direction is described by the time-dependent Hamiltonian
%
\begin{equation}
  H(t) = \frac{p^2}{2 M} + \alpha (t) x^2 + \beta (t) x^4 + \gamma (t) x,
  \label{eq:Hamiltonian}
\end{equation}
where $x$, $p$ are the position and momentum operators,  and $M$ the mass of the ion;  $\alpha(t)$, $\beta(t)$ and $\gamma(t)$ are in principle time-dependent coefficients.
%The Hamiltonian in Eq. \eqref{eq:Hamiltonian} can be implemented with segmented Paul traps where the values of the control parameters are set by tuning the voltages applied to the trap electrodes \cite{}.

%%%%%%%%%%%%%%%%%%%%%%%%%%%%%%%%%%%%%%%%%%%%%%%%%

The trapping potential $V_{t}(x) = \alpha (t) x^2 + \beta (t) x^4 + \gamma (t) x$ is a double well potential when $\alpha(t) < 0 $ and $\beta(t) > 0$. The  term $\gamma(t) x$  corresponds to a homogeneous electric field
that
%$\mathcal{E}(t) = \frac{\gamma(t)}{e}$ ($e=$ charge of the electron),
induces an energy bias between the wells.
%In this paper $\gamma$ is positive so  the right well is at higher energy than the one in the left.
%, and then, hereafter the lowest (highest) energy well will be referred to as the left (right) well.
%The position of the minima of each well is found from $\frac{\partial V_{t}}{\partial x} = 0$.
In the symmetric potential ($\gamma = 0$) the minima are  at $x_{0,\pm} = \pm \sqrt{-{\alpha}/{(2 \beta)}}$. We consider
%values of
the bias small enough so that the shift  of the minima
%with respect to the symmetric case
depends linearly on $\gamma$. The positions of the minima for a non-zero small bias are \cite{Martinez-Garaot2015}
%
%\begin{equation}
$
x_{0,\pm} \approx \pm \sqrt{-{\alpha}/({2 \beta})} + {\gamma}/({4 \alpha}),
$
%  \label{FIn the followingeq:PositionsMinima}
%\end{equation}
%
valid when  %\cite{Martinez-Garaot2015}
%
%\beq
%\label{sbc}
$
\abs{\gamma} \ll {4\sqrt{2}}\sqrt{-{\alpha^3}/{\beta}}/3,
$
%\eeq
%
which defines the small-bias regime.  In this regime  the energy difference between the wells is approximately $\Delta V_{t} = \gamma D$, where $D \equiv x_{0,+} - x_{0,-} = \sqrt{-{2\alpha}/{\beta}}$ is the distance between the minima.
%In the following t
The parameters for the initial double well will be chosen within this regime. The effective frequency $\omega_{eff,\pm}$ of each  well,
%
%\begin{equation}
%$
%\omega_{eff,\pm} \equiv \sqrt{\frac{1}{M}\left(\frac{\partial^2 V_{t}}{\partial x^2}\right)_{x=x_{0,\pm}}},
%$
%  \label{eq:EffevtiveFreq}
%\end{equation}
%
%has approximately the same value for both wells
in the small-bias regime is  $\omega_{eff,\pm} \approx \Omega= 2 \sqrt{{-\alpha}/{M}}$.

When $\alpha > 0$ and $\beta = 0$ the trapping potential is harmonic with angular frequency $\omega = \sqrt{{2\alpha}/{M}}$.
Multiplexing consists on driving the system from the double well configuration to the harmonic trap configuration so that the initial eigenstates are dynamically mapped onto the final ones. For simplicity we shall
%only consider changes in  $\alpha(t)$ and $\beta(t)$
keep $\gamma(t)$ fixed,
$\gamma(t) = \gamma$.
%The effect of the linear term $\gamma x$ in the harmonic trap is a displacement of the minimum  and a homogeneous $-{\gamma}/({4 \alpha^2)}$ shift of all the levels.
% that is unphysical.
The boundary conditions  in a multiplexing operation are $\alpha_0 <0$, $\beta_0 > 0$ for the initial values and  $\alpha_f >0$, $\beta_f = 0$ for the final values,
\begin{equation}
    \begin{split}
        V_{t=0}\left(x\right) &= \alpha_0 x^2 + \beta_0 x^4 + \gamma x,
        \\
        V_{t=t_f}\left(x\right) &= \alpha_f \left(x - x_{eq} \right)^2 -{\gamma}/({4 \alpha_f^2}),
    \end{split}
    \label{eq:InitialFinalPotentials}
\end{equation}
with $x_{eq}\equiv -{\gamma}/({2 \alpha_f})$.
We shall also impose that the frequency of the final harmonic trap is equal to the  frequency of the initial wells
%Since in the harmonic trap configuration the frequency is  $\omega = \sqrt{{2\alpha}/{M}}$, the final value of the quadratic term has to be
so $\alpha_f = 2 \abs{\alpha_0}$.
If the evolution is adiabatic, the lowest state of the upper well  ($n$th state globally) will become
the $n$th Fock  excited state $\ket{n}$ of the final harmonic potential.
%In order to have the ground state of the highest energy well as the $n$th excited state of the double well Hamiltonian the bias has to be large enough so there are exactly $n$ eigenstates in the lowest energy well below the ground state of the highest energy well.
If the wells are deep enough, in the left (right) well there is a set of harmonic eigenstates $\ket{n_L}$ ($\ket{n_R}$) with energies $E_{n_L} = \hbar \Omega_0 (n_L +1/2)$ $\left(E_{n_R} = \hbar \Omega_0 (n_R +1/2) + \Delta V_{t}\right)$, where $\Omega_0 \equiv 2 \sqrt{{-\alpha_0}/{m}}$. We need the initial ground state of the right well, $\ket{0_R}$, to be   the $n$th excited state of the whole system, so the inequality $E_{(n-1)_L} < E_{0_R} < E_{n_L}$ must be satisfied,
%  Thus, the bias to prepare the $n$th Fock state must satisfy
%
\begin{equation}
n - 1 < {\gamma D_0}/{(\hbar \Omega_0)} < n,
\label{eq:BiasCondition}
\end{equation}
where $D_0 \equiv D(\alpha_0,\beta_0)$. The ratio ${D}/{\Omega}=\sqrt{{M}/({2\beta})}$ only depends on $\beta$ so a change of $\alpha$ within the small
bias regime for constant $\beta$ does not modify this state ordering.
%If the equality $\frac{\gamma D_0}{\hbar \Omega_0} = n$ is satisfied there is going to be a degeneracy between the ground state of the highest energy well and the $n$th excited state of the lowest energy well. To avoid  energy degeneracy both with the $(n-1)$th and $n$th excited states of the lowest energy well
In our simulations we choose the value $\gamma = (n - {1}/{2}) {\hbar \Omega_0}/{D_0}$ for the bias. The small bias condition %\eqref{sbc}
and Eq. \eqref{eq:BiasCondition}  provide an upper bound for the highest Fock state that can be prepared with specific initial values of the control parameters $\alpha_0$ and $\beta_0$,
%
%
%\begin{equation}
$
  n \ll {4}\sqrt{-{M \alpha_0^3}/({\hbar^2\beta_0^2})}/3.
$
%  \label{eq:MaxNumberOfEigenstates}
%\end{equation}
%
To design  the driving of the control parameters, a straightforward approach would be an adiabatic evolution, for example a linear ramp protocol
along a large  run-time. Long times, however, are inadequate for many applications  and give rise  to decoherence.
Shortcuts to adiabaticity  \cite{Torrontegui2013,Guery-Odelin2019} stand out as a practical, faster option.
%
%\begin{equation}
%  \lim_{\alpha \to \alpha_0} \beta(\alpha) = A = \beta_0
%\end{equation}

%\begin{equation}
%  \lim_{\alpha \to \alpha_f} \beta(\alpha) = A + B = \beta_f
%\end{equation}
%
%
%
%
%
%
%\section{Design of the process}

{\it Design of the process.}
%\label{Section:Shortcuts}
%
%
%
%
%
%
Shortcuts to adiabaticity \cite{Torrontegui2013,Guery-Odelin2019} are a family of methods which speed up adiabatic processes  to get the same final populations or states in shorter times. Shortcuts have been applied for many different systems and operations and can be adapted to be robust against implementation errors and noise \cite{Guery-Odelin2019}.

%%%%%%%%%%%%%%%%%%%%%%%%%%%%%%%%%%%%%%%%%%%%%%%%%%%%%%%%%%%%%%%%%%%%%
%\begin{figure}
%%  \includegraphics[width = 0.75\linewidth]{BetaParametrization.eps}
%  \includegraphics[width = 0.75\linewidth]{PLOT_PROTOCOLS.eps}
  %
%\caption{Control parameter $\beta$ chosen to design the time dependence of $\alpha$, see Eq. (\ref{eq:S_function}).
%(b) $\alpha(t)$ for a linear ramp protocol (dashed-dotted, blue line); the Local Adiabatic protocol (short-dashed, red line);  and the FAQUAD protocol (solid, black line).
%$\alpha_0 = -4.7$ pN/m, $\alpha_f = - 2 \alpha_0 = 9.4$ pN/m, $\beta_0 = 0.052$ N/m$^3$, $\beta_f = 0$, $\gamma = 0.97$ zN, $\epsilon = 1$ pN/m, $\kappa  = 100/(\alpha_0-\alpha_f) = - 7.092$ m/pN.}
%\label{fig:Beta}
%\end{figure}
%%%%%%%%%%%%%%%%%%%%%%%%%%%%%%%%%%%%%%%%%%%%%%%%%%%%%%%%%%%%%

\begin{figure*}
  \centering
  \includegraphics[width =\linewidth]{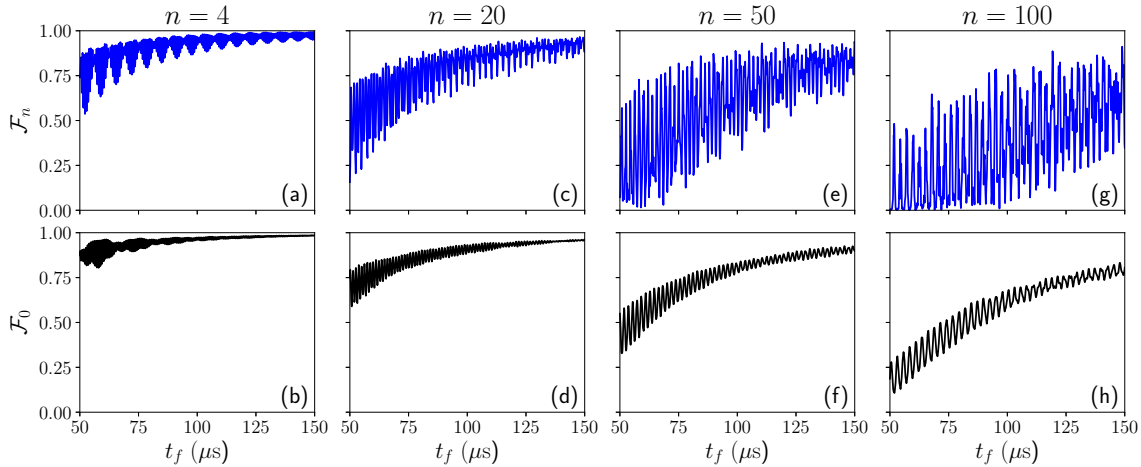}
  \caption{Preparation of Fock states $\ket{n}$ and superposition states  $({\ket{0}+e^{i\varphi}\ket{n}})/{\sqrt{2}}$ using FAQUAD. The different columns show the results for the different values of $n$. Upper row:  fidelity of the stated evolved by FAQUAD
  from the $n$th excited of the double well with respect to $\ket{n}$, the $n$th Fock state of the harmonic trap. Lower row: fidelity of the state evolved from the ground state with respect to $\ket{0}$ using the FAQUAD trap deformation designed to get $\ket{n}$.  $\alpha_0 = -4.7$ pN/m, $\alpha_f = - 2 \alpha_0 = 9.4$ pN/m, $\beta_0 = 0.052$ N/m$^3$, $\beta_f = 0$, $M(^{9}$Be$^{+}) = 9.012$ a.u, $\epsilon = 1$ pN/m, and $\kappa  = 100/(\alpha_0-\alpha_f) = - 7.092$ m/pN.}
  \label{fig:Fidelities_superposition}
\end{figure*}

Among the different STA techniques available, Fast quasiadiabatic dynamics (FAQUAD) \cite{Martinez-Garaot2015a} is
well suited to our current objective.
%t have been shown to accelerate adiabatic dynamics are not easily applied in the present scenario. Namely,
Invariants-based inverse engineering \cite{Chen2011} requires explicit knowledge of a dynamical invariant of the Hamiltonian,
which is not available here, and Fast-Forward driving \cite{Masuda2008,Torrontegui2012} produces potentials with singularities due to  the nodes of the wave function \cite{Garaot2016ArbDriving}, which can be problematic with highly excited states. FAQUAD reduces the diabatic transitions between the states of the Hamiltonian by making the adiabaticity criterion constant during  the process. For a time-dependent Hamiltonian that depends on a single control parameter $H(t) = H\left[ \lambda(t)\right]$ such that $\lambda(t)$ is a monotonous function in the $\left[0,t_f\right]$ interval, the adiabaticity criterion to avoid transitions between the instantaneous eigenstates $\ket{n(\lambda)}$ and $\ket{m(\lambda)}$ is \cite{Schiff1968}
%
%\begin{equation}
%  \pm \hbar \dot{\lambda} \; \abs{\frac{\braket{n(\lambda)}{\frac{d}{d\lambda}m(\lambda)}}{E_n(\lambda)-E_m(\lambda)}} = c \ll 1,
%  \label{eq:ad_criteria}
%\end{equation}
%
\begin{equation}
  \hbar\dot{\lambda} \; \abs{\frac{\matrixel{n(\lambda)}{dH/d\lambda}{m(\lambda)}}{\left[E_n(\lambda)-E_m(\lambda)\right]^2}} = c \ll 1,
  \label{eq:ad_criteria_practical}
\end{equation}
where $E_n(\lambda)$($E_m(\lambda)$) are the instantaneous eigenenergies and the dot stands for time derivative.
%The $+$ ($-$) sign in Eq. \eqref{eq:ad_criteria} is taken when $\lambda(t)$ is a strictly increasing (decreasing) function of time. Hereafter  the $+$ sign is  taken since $\lambda(t)$ will increase with time.
%In Eq. \eqref{eq:ad_criteria} a derivative of one of the eigenstates with respect to the parameter $\alpha$ appears which is cumbersome to compute specially when the eigenstates are obtained numerically which forces to choose a consistent phase for them.
%Equation  \eqref{eq:ad_criteria} can also be written as
%

%
%because of the identity $\matrixel{n(\lambda)}{\frac{d}{d\lambda}H(\lambda)}{m(\lambda)}= (E_m-E_n)\braket{n(\lambda)}{\frac{d}{d\lambda}m(\lambda)}$ \cite{Schiff1968}.
FAQUAD imposes a constant $c$, so Eq. \eqref{eq:ad_criteria_practical} becomes  a differential equation for   $\lambda(t)$. The value of $c$ is determined  by the boundary conditions
$\lambda(0)$ and $\lambda(t_f)$.
Equation \eqref{eq:ad_criteria_practical} implies that the control parameter evolves
more slowly  when
% the ratio $\abs{\frac{\matrixel{n(\lambda)}{\frac{d}{d\lambda}H(\lambda)}{m(\lambda)}}{\left[E_n(\lambda)-E_m(\lambda)\right]^2}}$ has  larger  values, i.e. when
the Hamiltonian changes rapidly with the control parameter and/or near avoided crossings.

We eliminate one degree of freedom in Eq. (\ref{eq:Hamiltonian}) by taking  $\alpha$ as the master control parameter ($\alpha = \lambda$) and making
%$\beta$ to be $\alpha$-dependent, i.e.
$\beta = \beta(\alpha)$.
%Now, the derivative operator in Eqs. \eqref{eq:ad_criteria} and \eqref{eq:ad_criteria_practical} becomes $\frac{d}{d\alpha} = \frac{\partial}{\partial \alpha} + \frac{d\beta}{d\alpha}\frac{\partial}{\partial \beta}$ by the chain rule.
%Then  Eq. \eqref{eq:ad_criteria_practical} becomes an ordinary differential equation for $\alpha$
% since all the terms in Eq. \eqref{eq:ad_criteria_practical} now depend only on that parameter.
%With the Hamiltonian in Eq. \eqref{eq:Hamiltonian} we get
%
%\begin{equation}
%$
%\frac{dH}{d\alpha} = x^2 + \frac{d\beta}{d\alpha} x^4.
%$
%\end{equation}
%
%To design the control parameters of the Hamiltonian \eqref{eq:Hamiltonian}, first,
Eq. \eqref{eq:ad_criteria_practical} has to be solved  with the boundary conditions for $\alpha$ and $\beta$.
%$\alpha(0) = \alpha_0$, $\alpha(t_f) = \alpha_f$, $\beta(\alpha_0) = \beta_0$ and $\beta(\alpha_f) = \beta_f$.
To choose $\beta(\alpha)$ we consider that the largest possible values of $\beta$ should hold while $\alpha$ changes sign so that the
levels in the intermediate quartic well are not too close.
% to each other and promote transitions.
A simple choice is to keep   $\beta\approx\beta_0$
constant until $\alpha>0$ increases and the quadratic part dominates. Then we can let $\beta$ drop to zero without any significant effect.
While $\beta$ is constant the energy difference between the wells in units of the instantaneous motional quantum
%
%\begin{equation}
$
N_q = {\gamma D }/({\hbar \Omega}) = {\gamma} \sqrt{{M}/{(2\beta)}}/\hbar,
$
%\label{eq:depth}
%\end{equation}
%
is  constant. We choose for $\beta$ the form
%
%\begin{equation}
$
  \beta(\alpha) = a + b\;S\left[\kappa(\alpha - \epsilon)\right],
$
%  \label{eq:S_function}
%\end{equation}
%
where $S(x) =(1+e^{-x})^{-1}$  is the (sigmoid) logistic function.
% \cite{Mitchell1997}.
%, and the values of $a$ and $b$ are determined by the boundary conditions of $\beta$.
%The sigmoid function  jumps from  $0$ for $x$ negative to $1$ for $x$ positive
%around $x = 0$, see Fig. \ref{fig:Beta}. {With this parametrization the value of $\beta$ is approximately constant for $\alpha<0$ and $\alpha>0$ with a sudden jump around $\alpha\approx 0$}
%
$\epsilon$ and $\kappa$
%in Eq. \eqref{eq:S_function}
set the position and the width of the region where the parameter $\beta$ ramps from its initial to the final value.
%one, respectively.
We choose $\epsilon>0$ so that $\beta\approx \beta_0$ around $\alpha=0$.
%, see Fig. \ref{fig:Beta}.
A larger $\kappa$ implies a narrower jump. When $\kappa \gg max \{ \left(\abs{\alpha_0}+\epsilon\right)^{-1},\left(\abs{\alpha_f}-\epsilon\right)^{-1}\}$ the ramp of $\beta$ is narrow enough so that when $\alpha$ goes to $\alpha_0$ ($\alpha_f$) $\beta$ goes asymptotically to $a$ ($a + b$) and then the boundary conditions  demand that $a = -b = \beta_0$.

We choose $\alpha_0 = - 4.7$ pN/m, $\alpha_f = 2 \abs{\alpha_0} = 9.4$ pN/m, $\beta_0 = 0.052$ N/m$^3$, and $^9$Be$^+$ ions in  the numerical simulations.
%The inequality in Eq. \eqref{eq:MaxNumberOfEigenstates} is sufficiently satisfied for the $n=100$ Fock states, which is the largest one studied here, since $n \ll \frac{4}{3}\sqrt{-\frac{M \alpha_0^3}{\hbar^2\beta_0^2}} \approx 3\times 10^5$.
%We shall also impose that the frequency of the final harmonic trap is equal to the  frequency of the initial wells
%Since in the harmonic trap configuration the frequency is  $\omega = \sqrt{{2\alpha}/{M}}$, the final value of the quadratic term has to be
%so .
With the chosen $\alpha_0$, $\beta_0$ and the mass of $^9$Be$^+$,  $\Omega_0 = 2\pi\,\times\,5.6$ MHz and  $D_0 = 13.45$ $\mu$m. For the ground state of the highest energy well to be the $n$th excited state of the full Hamiltonian, the bias is chosen as $\gamma = (n-0.5) {\hbar \Omega_0}/{D_0}$.

Avoided level crossings occur at $\alpha<0$, near $\alpha=0$
%, see Fig. \ref{fig:Energies},
in a critical region
where the small bias condition fails and the double well becomes a single  quartic well.
% to which the levels that were originally in the upper well are transferred.
% As can be seen in Eq. \eqref{eq:depth} the number of eigenstates in the lower well depends only on $\beta$ so it is convenient to keep its value constant during the part of the process in which the trap is a double well ($\alpha<0$).
%The logistic function is suitable for this purpose since the value of $\beta$ will be nearly constant for $\alpha \lesssim 0$.
%The region near $\alpha=0$ is important because the double barrier disappears and all levels get close to each other as can be seen in . There,
The gap between the eigenstates near $\alpha=0$ is approximately proportional to $\beta^{1/3}$
%which can be proven using the \textit{WKB} approximation for a purely quartic potential
\cite{Vranicar2000}. Thus, at $\alpha \approx 0$, $\beta$ should be as large as possible within experimental constraints.
%this is not always possible due to experimental limitations \textcolor{blue}{\textbf{(¿Buscar cita?)}}

%%%%%%%%%%%%%%%%%%%%%%%%%%%%%%%%%%%%%%%%%%%%%%%%%%%%%%%%
\begin{figure}
  \centering
  \includegraphics[width = 0.9\linewidth]{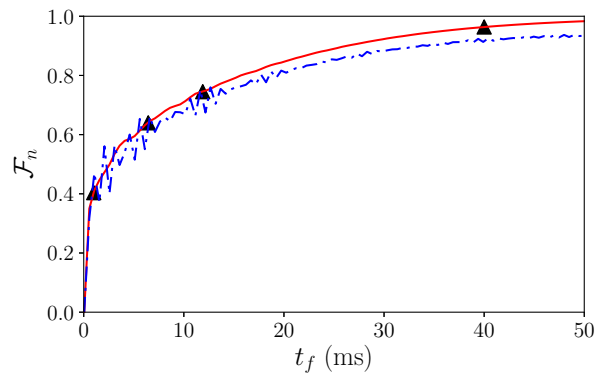}
  \caption{Fidelity vs. final time for the creation of Fock states using a linear ramp of the control parameter $\alpha$: triangles, $n = 20$; solid red line, $n = 50$ and dashed blue line $n = 100$. Same parameters as in Fig. 2.}
%  $\alpha_0 = -4.7$ pN/m, $\alpha_f = - 2 \alpha_0 = 9.4$ pN/m, $\beta_0 = 0.052$ N/m$^3$, $\beta_f = 0$, $M(^{9}$Be$^{+}) = %9.012$ a.u, $\epsilon = 1$ pN/m, $\kappa  = 100/(\alpha_0-\alpha_f) = - 7.092$ m/pN.}
\label{fig:Fidelities_LinearRamp}
\end{figure}
%%%%%%%%%%%%%%%%%%%%%%%%%%%%%%%%%%%%%%%%%%%%%%%%%%%

%Equation \eqref{eq:ad_criteria_practical} gives a good adiabaticity criterion for two-level systems where the
%application of FAQUAD is straightforward \cite{Martinez-Garaot2015a}. However,
In our multilevel
scenario  we modify Eq. \eqref{eq:ad_criteria_practical}  to \cite{Palmero2019}
%
%\begin{equation}
$
 c = \hbar\dot{\lambda} \sum_{m \neq n}  \abs{\frac{\matrixel{n(\lambda)}{{dH/}{d\lambda}}{m(\lambda)}}{\left[E_n(\lambda)-E_m(\lambda)\right]^2}},
$
%  \label{eq:FAQ_SUM}
%\end{equation}
%
%which gives the FAQUAD equation to design $\alpha(t)$ for $c$ constant. The sum is in principle for all the eigenstates of the Hamiltonian different from $\ket{n}$ but only a few of them will be significant.
taking only the four closest eigenstates (two from below and two from above) of the relevant state in the sum. Note the
shorthand notation $|n\ra\equiv |n(\lambda_f)\ra$ for the eigenstates of the final harmonic oscillator.
%(see Appendix \ref{append:n4}).

%
%
%
%\section{Results}
%\label{Section:Results}
%
%
%
%%%%%%%%%%%%%%%%%%%%%%%%%%%%%%%%%%%%%%%%%%%%%%%%%%%%%%%%%%%%%%%%%%%%%%%%

%
%
%\subsection{Time evolution}
%
{\it Results.}
We have numerically solved the time-dependent Schr\"odinger equation for the Hamiltonian  \eqref{eq:Hamiltonian}.
We  compare the performance of the protocols designed using FAQUAD with a linear ramp of the control parameter $\alpha(t)$, using the same $\beta(\alpha)$ as for FAQUAD.
%

%\subsection{Fock State creation}
%{\it Fock state creation.}
%
%
The upper  row of Fig. \ref{fig:Fidelities_superposition} shows the results of the multiplexing step ((c) in Fig. \ref{fig:ProtocolDiagram}) for different $n$. The fidelity is  $\mathcal{F}_n = \abs{\braket{n}{\psi_n^{F}}}$,  where $\ket{\psi_n^{F}}$ is the final state after FAQUAD evolution.
%The panels correspond to: (a) $n = 4$, (c) $n= 20$, (e) $n = 50$, and (g) $n= 100$.
The fidelity if $\alpha(t)$ follows a linear ramp is depicted in Fig. \ref{fig:Fidelities_LinearRamp}.
%As can be seen in Fig. \ref{fig:Fidelities_superposition} and Fig. \ref{fig:Fidelities_LinearRamp},
%The  performance of FAQUAD and the linear ramp are quite  different.
FAQUAD attains fidelities above $\mathcal{F} = 0.9$ for final times of less than $100\;\mu\mathrm{s}$,
%(for the $n = 4 $ Fock state),
while the linear ramp needs evolution times up to $50\;m\mathrm{s}$ for similar fidelities. In Fig. \ref{fig:Fidelities_superposition} (upper panels)  the maximum fidelities  for similar final times decrease and  the width of the fidelity oscillations increases for larger $n$. Both effects can be mitigated using a local adiabatic approach, see the final discussion.  Nevertheless, for the studied final times, fidelities above $\mathcal{F} = 0.9$ for  $n= 100$ can be reached for specific values of $t_f$. In Ref. \cite{McCormick2019} a table shows the final times required to create each Fock state by combining Rabi pulses.
% also with a Berilium ion.
For $n = 4$ only 38 $\mu$s are needed, but for  $n = 100$ the total time grows to 335 $\mu$s, even though higher order sidebands were applied. In comparison, even if the times needed are orders of magnitude larger, the remarkable stability
of the fidelity curve with respect to $n$ is noteworthy for the linear ramp in Fig. \ref{fig:Fidelities_LinearRamp}. This stability
of a trap deformation method  also holds, although somewhat weakened, in the upper edge of the fidelity curve using FAQUAD, which may be useful assuming that scans on final time can be made.   %Moreover, it should easily allow for even higher Fock state creation \footnote{As the system size grows, smaller energy gaps increase the computational requirements for the time evolution, but obtaining the FAQUAD and LA protocols do not require so many resources}.

The explanation for the decreasing fidelities as the process aims at higher Fock states is that the nearest energy levels get closer, and transitions
among more and more levels occur making the interference pattern, inherent in FAQUAD \cite{Martinez-Garaot2015}, more  complex.
%.  For $n = 4$, the oscillations of the fidelity vs. final time are very regular, and can be understood in terms of a few transitions,
%(see Appendix \ref{append:n4} for a detailed study),
%whereas for higher Fock states
%these oscillations become increasingly complex.

%
%
{\it{Preparing superpositions.}}
\begin{figure}[t]
  \centering
  \includegraphics[width = 1.01\linewidth]{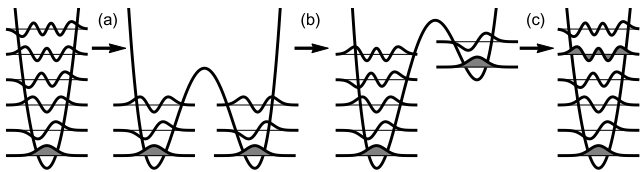}
  \caption{Scheme to prepare superpositions of the ground state and a Fock state of the harmonic oscillator,  $({\ket{0}+e^{i\varphi}\ket{n}})/{\sqrt{2}}$. (a) Splitting of the ground state; (b) Biasing; (c) Merging
the two wells  into a harmonic trap. The shaded wave functions are the ideal initial and target states.}
  \label{fig:SuperpositionProtocolDiagram}
\end{figure}
%
%%%%%%%%%%%%%%%%
\begin{figure}[t!]
\centering
\vspace*{.0cm}
  \includegraphics[width = 1.\linewidth]{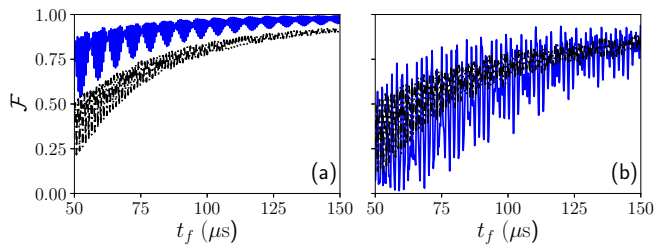}
  \caption{Comparison of the fidelities obtained after performing the multiplexing process using FAQUAD (solid blue lines) and LA method for: (a) $n=4$ and (b) $n=50$. Same parameters as in Fig. 2}
% $\alpha_0 = -4.7$ pN/m, $\alpha_f = - 2 \alpha_0 = 9.4$ pN/m, $\beta_0 = 0.052$ N/m$^3$, $\beta_f = 0$, $M(^{9}$Be$^{+}) = %9.012$ a.u., $\gamma = 0.97$ zN, $\epsilon = 1$ pN/m, $\kappa  = 100/(\alpha_0-\alpha_f) = - 7.092$ m/pN.}
  \label{fig:FAQUAD_AND_LA}
\end{figure}
%%%%%%%%%%%%%%%%
%
%%%%%%%%%%%%%%%%%%%%%%%%%%%%%%%%%%%%%%%%%%%%%%%%%%%%%%%%%%%%%%%%%%%%%%
%\begin{figure}
%  \centering
%  \includegraphics[width = 0.9\linewidth]{Figure_Wrong_Eigenstate.eps}
%  \caption{Fidelity vs. final time for the creation of the $n = 90$ ($\gamma = 89.5 \frac{\hbar \Omega_0}{D_0}$) Fock state using the protocol designed for the $n = 100$. $\alpha_0 = -4.7$ pN/m, $\alpha_f = - 2 \alpha_0 = 9.4$ pN/m, $\beta_0 = 0.052$ N/m$^3$, $\beta_f = 0$, $M(^{9}$Be$^{+}) = 9.012$ a.u, $\epsilon = 1$ pN/m, $\kappa  = 100/(\alpha_0-\alpha_f) = - 7.092$ m/pN.}
%\label{fig:EvolutionOfWrongEigenstate}
%\end{figure}
%%%%%%%%%%%%%%%%%%%%%%%%%%%%%%%%%%%%%%%%%%%%%%%%%%%%%%%%%%%%%%%%%%%%%%
%
The protocols studied so far are for a pure Fock state preparation. However, they also allow us to prepare states  $|\psi_{\varphi}\rangle = \left(|0\rangle+e^{i\varphi}|n\rangle\right)/\sqrt{2}$ up to a relative phase $\varphi$.

%We have tested what happens when we design the multiplexing process to prepare a certain Fock state $\ket{n}$ but, instead, we use this protocol to prepare another Fock state $\ket{n'}$ ($n\neq n'$). For example, consider the scenario where we design the multiplexing process using FAQUAD to prepare the $n=100$ Fock but we set the bias (second step in Fig. \ref{fig:ProtocolDiagram}) to prepare the $n'= 90$ Fock state. We want to know if the final state of the process will be the 90th Fock state. Figure \ref{fig:EvolutionOfWrongEigenstate} demonstrates that in this case the fidelity does not deteriorate significantly with respect Fig. \ref{fig:Fidelities_superposition}(g). In Fig. \ref{fig:EvolutionOfWrongEigenstate} we show that a protocol that was initially designed for a particular Fock state can be used to prepare other Fock states just by changing the initial bias of the trap.

%This robustness gives us hope that, even if the protocols are exclusively designed to create Fock states, they may also work for superposition states.
A modification of the sequence in Fig. \ref{fig:ProtocolDiagram} leads to superposition states, see Fig.  \ref{fig:SuperpositionProtocolDiagram}.
%We first split the ground state wave function of a harmonic trap into the symmetric superposition of states of the double well (Fig. \ref{fig:SuperpositionProtocolDiagram} step (a)), then the double well is biased to create the necessary energy difference between the wave functions (step (b)) and finally (step (c)) the wave functions are merged back into a harmonic trap. In this paper we report results of the simulation of step (c). Because of the linearity of the Schr\"odiner equation, we can simulate independently the evolution of the ground state $\ket{\psi_0}$ and the $n$th excited state of the biased double well into the final harmonic trap.
The success of the protocol (step (c)) is measured with the fidelity $\mathcal{F}_n$ to reach $\ket{n}$ starting in the $n$th excited state of the double well, and the fidelity $\mathcal{F}_0(n)$ to reach $\ket{0}$ starting in the ground state while using the deformation devised to reach $\ket{n}$. (The average $(\mathcal{F}_0(n)+\mathcal{F}_n)/2$ is the maximal fidelity with respect to the states labeled by $\varphi$).
%evolving 4Nies  and , being the fidelities of the initial ground state and  evolving into the ground state $\ket{0}$ and Fock state $\ket{n}$ of the harmonic trap respectively.
%When the Fidelities are close to one, the final state approaches $\frac{1}{\sqrt{2}}\left(|0\rangle+e^{i\varphi}|n\rangle\right)$.
The upper row of Fig. \ref{fig:Fidelities_superposition} pictures $\mathcal{F}_n$ for  $n = 4, 20, 50, 100$
%for the $n$th Fock state panels (a), (c), (e), (g),
and the lower panels the corresponding $\mathcal{F}_0(n)$. The $\mathcal{F}_0(n)$ are remarkably close to the $\mathcal{F}_n$, which makes superpositions $|\psi_{\varphi}\rangle$ feasible with high fidelity.
%The fidelities for the ground state are slightly worse than for the $n$th Fock state, but we still reach maximum fidelities close to the latter.

In \cite{McCormick2019}   these superpositions were created via Rabi pulses
for measuring deviations from a nominal trap frequency. The maximum sensitivity was reached for the superposition of the ground and the 12th state.
%With the scheme of trap modifications, and with the fidelities showed here, it should be possible to reach much higher superposition states to a good fidelity, which can potentially lead to even better sensitivities.
%

%\section{Discusion}
%\label{Section:Conclusions}
{\it Discussion.}
We have proposed to
%fast multiplexing operations for vibrational states of trapped ions which, combined with demultiplexing and bias-inversion protocols, enable us to
prepare highly excited Fock states and superpositions  with the ground state in trapped ions using deformations
%of the trapping potential
between double and single wells. Since no Rabi pulses are involved, these protocols can be applied to different atomic species or particles.
%We have demonstrated the creation of up to the $n=100$ state of the final harmonic trap and the superposition of the ground and the 100th states.

A FAQUAD approach which distributes diabatic transitions homogeneously through all the process provides a significant speedup with respect to a linear ramp of the control parameter.
%Trap deformations based on the fast-forward approach \cite{Masuda2008,Torrontegui2012} were also considered in \cite{Garaot2016ArbDriving} to create Fock states
%but it implied potential divergences not found in the current proposal.
%. The current method requires however much simpler trap deformations.  requires intermediate potentials that diverge at the nodes of the instantaneous eigenstates. Truncated versions (i.e. without the divergences) can still be useful, however, as higher excited states have more nodes the intermediate potential will have more singularities so truncation will reduce the fidelity.
Methods similar in spirit to FAQUAD may also be applied \cite{Guery-Odelin2019}.
The Local Adiabatic (LA) method \cite{Roland2002} only uses the instantaneous energy gap between the eigenstates  to modulate the rate of change of the control parameter. Adapted to our multilevel scenario, we set
%
%\begin{equation}
$
c_{LA} = \hbar\dot{\lambda}_{LA} \sum_{m \neq n} {[E_n(\lambda)-E_m(\lambda)]^{-2}},
$
%  \label{eq:MONROE}
%
%\end{equation}
%
as a constant given by the boundary conditions,
%While in \cite{Roland2002} only the gap between the ground state and the first dynamically coupled excited is taken into account, we also include the gap with other states in parallel to the FAQUAD protocol.
and  parameterize $\beta(\alpha)$ as before.
%for the FAQUAD protocol so that the same energy levels are used.
We have compared the performance of FAQUAD against the LA method in Fig. \ref{fig:FAQUAD_AND_LA}. For small $n$  FAQUAD clearly outperforms the LA, but LA is more stable as $n$ increases, due to a lesser role of quantum
interferences \cite{Martinez-Garaot2015a}.
%as we grow $n$ it closes the gap, until $n = 100$, where they practically give equally high fidelities.

This paper demonstrates the potential of trap deformations  to control motional states. Future work  could be to find protocols for ion chains, and to make full use of the dimensionality of the parameter space \cite{Rezakhani2009}, reduced here to one for simplicity. The trap deformation in our model passes through a quartic potential well with close levels that plays the role of the bottleneck of the process speed. The search for  smooth, doable functional forms for the time dependence of the trap increasing the minimal gap, combined with  numerical optimization of the deformation is a worthwhile objective.

%------------------------------------------------------------------------------%
%             End of the Body of the paper                                     %
%------------------------------------------------------------------------------%
\section*{Acknowledgments}%
We acknowledge discussions with Joseba Alonso.
This work was supported by
the Basque Country Government (Grant No.
IT986-16) and PGC2018-101355-B-I00 (MCIU/AEI/FEDER,UE).
M. P. acknowledges support from the Singapore Ministry of Education, Singapore Academic Research Fund Tier-II (project MOE2018-T2-2-142).

\bibliography{Fock_Collection.bib}
\bibliographystyle{apsrev4-1}
\end{document}